\documentclass[superscriptaddress, reprint, amsmath, amssymb, aps, pra, floatfix]{revtex4-2}
\usepackage{graphicx} % Required for inserting images

\usepackage{xcolor} % for colored text
\usepackage{hyperref} % for hyperlinks
\usepackage{orcidlink} % for ORCID iDs
\hypersetup{
	colorlinks=true,
	linkcolor=blue,
	filecolor=magenta,
	urlcolor=magenta,
}

\definecolor{dgreen}{rgb}{0.0,0.5,0.0}
\definecolor{pink}{rgb}{1,0,0.9}

\usepackage[normalem]{ulem}

\begin{document}
\title{SU(4) Heisenberg model on the hyperhoneycomb lattice: Tensor network study}

\author{I.V. Lukin\orcidlink{0000-0002-8133-2829}}
\email{illya.lukin11@gmail.com}
\affiliation{Akhiezer Institute for Theoretical Physics, NSC KIPT, Akademichna 1, 61108 Kharkiv, Ukraine}
\affiliation{Haiqu, Inc., 95 Third Street, San Francisco, CA 94103, USA}

\author{A.G. Sotnikov\orcidlink{0000-0002-3632-4790}}
\email{a\_sotnikov@kipt.kharkov.ua}
\affiliation{Akhiezer Institute for Theoretical Physics, NSC KIPT, Akademichna 1, 61108 Kharkiv, Ukraine}
\affiliation{Education and Research Institute ``School of Physics and Technology'', Karazin Kharkiv National University, Svobody Square 4, 61022 Kharkiv, Ukraine}

\begin{abstract}
    We study the ground state of the SU(4) Heisenberg model on the hyperhoneycomb lattice using three-dimensional projected entangled pair states. We show that it is possible to compute physical observables for the ground states using loop expansions, which converge quickly on tree-like lattices. Our extrapolations to the limit of infinite bond dimensions point toward a gapless spin-liquid ground state.  
\end{abstract}
\date{\today}

\maketitle

\section{Introduction}
SU($N$) generalizations of the Heisenberg model have attracted considerable interest over the past two decades, driven by two complementary experimental directions. On one hand, ultracold alkaline-earth(-like) atoms (such as $^{173}$Yb or $^{87}$Sr) loaded into optical lattices provide a remarkably clean realization of SU($N$)-symmetric interactions: the nuclear spin decouples from the electronic angular momentum in the ground state, so that interactions become independent of the nuclear spin orientation, yielding an emergent SU($N$) symmetry with $2\leq N\leq10$ \cite{Gorshkov2010,Cazalilla2009,CazalillaRey2014}. 
The high degree of control over lattice geometry, dimensionality, and filling achievable in cold-atom experiments has motivated extensive theoretical and experimental studies of SU($N$) Heisenberg and Hubbard models as targets for quantum simulation \cite{Hermele2009, Taie2012, Scazza2014, Zhang2014, Pagano2014}.

On the other hand, SU($N$) and, more generally, Kugel-Khomskii-type spin-orbital models arise as effective low-energy descriptions of strongly correlated electron materials with active orbital degrees of freedom in addition to spin \cite{KugelKhomskii1982, Oles2009}. In transition-metal compounds with partially filled $d$-shells, strong on-site Coulomb repulsion combined with orbital degeneracy leads to superexchange Hamiltonians in which
spin and orbital operators are intertwined. In materials with strong spin-orbit coupling, spin and orbital degrees of freedom can instead be locked together into a single atomic multiplet. In particular, for $4d^1$ and $5d^1$ ions in ordered double perovskites, the resulting $j_{\rm eff}=3/2$ quadruplet on each site forms (approximately) the fundamental representation of SU(4), and the corresponding superexchange Hamiltonian acquires an enlarged SU(4) symmetry~\cite{ChenBalents2010, Romhanyi2017, Natori2018, YamadaSU42018,Jin2023}. This scenario has been proposed as relevant to candidate spin-orbital liquid materials such as Ba$_2$YMoO$_6$~\cite{Natori2018}, making such compounds natural solid-state platforms for studying SU(4) physics, including the possibility of exotic, symmetric quantum spin-liquid ground states.

In both settings---cold atoms and spin-orbital materials---the enlarged
symmetry suppresses long-range magnetic order more strongly than in the
SU(2) case \cite{Wu2013}, making SU($N$) models fertile ground for searching for novel
quantum disordered phases: chiral and non-chiral spin liquids,
valence-bond solids, and other exotic states without simple classical
analogues.

With respect to one-dimensional (1d) and two-dimensional (2d) lattice geometries, quasi-exact theoretical approaches, such as tensor network methods \cite{Cirac2021MPSPEPSReview, Orus2014PracticalIntro, Orus2019ComplexQuantumSystems, Banuls2023RouteMap, Schollwoeck2011DMRGAgeMPS}, have proven to be efficient in studies of the SU($N$)-symmetric Heisenberg models. In particular, tensor network states were employed to predict gapless and gapped spin liquid states, plaquette valence bond solids, dimerized states, and various magnetic orderings \cite{Corboz2012SimplexSolids, Corboz2013CompetingStatesSU3, Corboz2012SpinOrbitalLiquid, Nataf2016SU6Honeycomb, Bauer2012ThreeSublatticeSU3, Xu2023KagomeSU3Phase, Corboz2011SU4DimerizationSquare, LeeKawashima2018StarLatticeBLBQ}. 

Unfortunately, this success is hard to generalize to three-dimensional (3d) settings. The main problem is that it is very computationally costly to obtain expectation values of operators with three-dimensional tensor-network states. By default, the observables are computed with the contraction of 3d tensor networks by means of auxiliary 2d and 1d tensor network states \cite{VlaarCorboz2021Simulation3D, LukinSotnikov2024SingleLayer3D, LukinSotnikov2025SU4CubicDimerization}. This can be realized only for very moderate bond dimensions~$D$ (up to $D=7$ at this moment), which may be insufficient for systems with SU($N$) symmetry. 

Recently, Ref.~\cite{Dziarmaga2026MonteCarlo3DPEPS} proposed employing Monte Carlo sampling to compute observables with 3d tensor-network states. This method potentially has a lower computational cost, but it works only for finite systems with open boundary conditions. Yet another approach is to compute observables only approximately, with the help of belief propagation (Simple Update) environments \cite{JahromiOrus2019UniversalGPEPS, JahromiOrusPoilblancMila2020KagomeBreathing, JahromiOrus2020ThermalBosons3D, JahromiYarlooOrus2021Kitaev3D, TindallFishman2023GaugingBP, TindallFishmanStoudenmireSels2024EagleKickedIsing, TindallMelloFishmanStoudenmireSels2026DisorderedTN}. This approach does not have a clear error estimation, but it was shown to work rather well for gapped systems.  In this study, we explore another recently proposed method to compute observables---loop expansions on top of belief propagation \cite{EvenblyPancottiMilstedGrayChan2026LoopSeries, ChertkovChernyak2006LoopCalculus, ChertkovChernyak2006LoopSeries} (see also a recent alternative version---loop cluster expansion \cite{ParkGrayChan2025InfluenceFunctionalBP, MidhaZhang2025BeyondBP, MidhaSommersTindallAbanin2026RigorousBP, GrayParkEvenblyPancottiKjonstadChan2025LoopClusterExpansions} and a recent related proposal of generalized belief propagation~\cite{TindallSommersKappen2026GBP}). The loop expansion method builds a systematic expansion in loops on top of the belief propagation observables, which allows one to estimate errors and systematically improve the observables toward their true values. It was recently applied to the finite temperature frustrated magnets on the ruby lattice \cite{MelloStoudenmireTindall2026RubyFiniteT}. 

The loop expansion is expected to converge quickly on the tree-like lattices with large loops. In three dimensions, the prime fitting lattice candidates are trivalent lattices like hyperhoneycomb, stripy-hyperhoneycomb, hyperoctagon, as well as other harmonic  honeycomb lattices. Interestingly, the SU(4) Heisenberg model on the hyperhoneycomb lattice has already been studied in the literature and was proposed to host a gapless spin liquid ground state \cite{Natori2018}. In this study, we use tensor network methods with loop expansions to provide additional evidence that the SU($4$) Heisenberg model on the hyperhoneycomb lattice has a spin-liquid ground state and to simultaneously establish loop expansions as a promising approach to 3d tree-like lattices. 

The paper is organized as follows: in Sec.~\ref{sec:model}, we introduce the hyperhoneycomb lattice structure and the SU($4$)-symmetric Heisenberg model; in Sec.~\ref{sec:method}, we discuss the tensor network methods applied in the study; in Sec.~\ref{sec:results}, we present our main results. Finally, in Sec.~\ref{sec:conclusion}, we summarize our observations and give an outlook for future work. In addition, Appendix~\ref{appendix:loop} provides a technical discussion of the treatment of disconnected graphs in loop expansion, while Appendix~\ref{appendix:comparison} contains a comparison of loop expansions with more recent loop cluster expansions.

\section{Model}\label{sec:model}
In this work, we focus on the antiferromagnetic SU(4)-symmetric Heisenberg model on the hyperhoneycomb lattice. On every site of the lattice, we place a local Hilbert space of dimension $d_\alpha=4$, carrying the fundamental representation of the SU(4) group. The Hamiltonian of this model has the following form:
\begin{equation}\label{eq:model}
    H = \sum_{\langle ij \rangle}  P_{i,j},
\end{equation}
where $\langle ij \rangle$ denotes a summation over the nearest-neighbor pairs of sites, and $P_{ij}$ is the permutation operator, which acts on the two local basis states on sites $i$ and $j$ as follows: $P_{ij} |\alpha_i \beta_j\rangle = |\beta_i \alpha_j \rangle$. 

\begin{figure}
    %\centering
    \includegraphics[width=\linewidth]{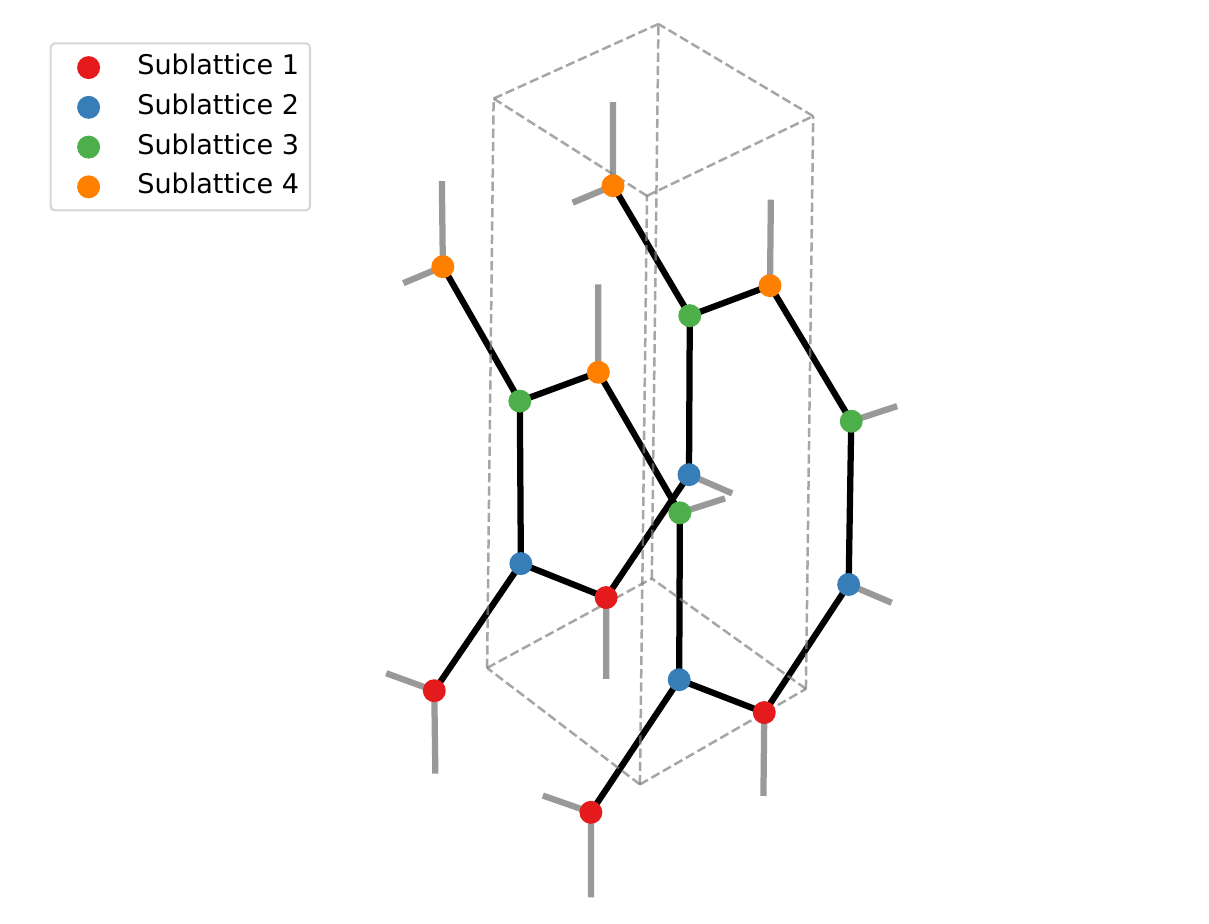}
    %{hyperhoneycomb_3d.pdf}
    \caption{Hyperhoneycomb lattice with a $1\times1\times1$ unit cell (the four unit cells are shown). The unit cell consists of four sites indicated by different colors and forming separate sublattices. Note that every site is connected to the sites from three other sublattices.}
    \label{fig:1}
\end{figure}

The unit cell of the hyperhoneycomb lattice is shown in Fig.~\ref{fig:1} and consists of four different sites and six different bonds, while all sites in this lattice are trivalent. The minimal loop in the system consists of $10$ different sites. Note that bonds pointing vertically (in $z$ direction) are not geometrically equivalent to other bonds. 

This model was already studied with a variational Monte Carlo approach in relation to spin-orbital models and was predicted to host a gapless spin-liquid phase~\cite{Natori2018}. 

%The SU(4) Heisenberg model on the hyperhoneycomb lattice is relevant to the physics of ultracold atoms and also to the physics of certain  $4d^{1}$ and $5d^{1}$ materials with strong spin-orbit coupling \cite{Natori2018}. In the latter case, the SU(4) symmetry is only approximate, but it is still relevant as a first approximation. \fixme{Ref} studied the model in the context of $d^{1}$ materials with the variational Monte-Carlo approach and argued that the ground state is a gapless spin liquid. The goal of this study is to confirm this suggestion with tensor-network methodology.  

\section{Methods}\label{sec:method}
We rely on the 3d tensor network variational ansatz---projected entangled pair state (PEPS) \cite{Cirac2021MPSPEPSReview, Orus2014PracticalIntro, Bruognolo2021BeginnersiPEPS}. The ansatz consists of rank-4 tensors $T^{p}_{ijk}$ placed on the sites of the infinite-size hyperhoneycomb lattice. The physical index $p$ takes values in the local $4$-dimensional Hilbert space on the lattice site. Auxiliary (or virtual) indices $i$, $j$, and $k$ run over the auxiliary vector space of dimension~$D$ (bond dimension) and correspond to the edges of the lattice. As the lattice edge connects two sites, the index corresponding to this edge is shared between the two tensors and is summed over. 

Note that the tensors on different lattice sites should not be identical. In view of this, we generally choose a periodically repeating unit cell of the lattice and place different tensors on the sites in this unit cell. The tensors corresponding to sites in other unit cells are obtained by periodicity. With this construction, it is possible to define the wave function directly in the infinite-lattice limit and with a well-defined pattern of translational symmetry breaking. The wave function is constructed in a way to be an accurate approximation for the ground state of the model. The bond dimension~$D$ is the main parameter that controls the accuracy of the calculation. It is expected that the ansatz becomes exact in the limit $D \to \infty$. In practice, one works with finite and relatively small values of $D$ ($D \lesssim 20$) and then extrapolates the obtained results to the limit $D=\infty$. 

After defining the wave function ansatz itself, the next goal is to optimize its parameters to approximate the many-body ground state. This can be achieved by means of the Simple Update method \cite{Jiang2008SimpleUpdate}. The Simple Update method simulates PEPS ansatz evolution in imaginary time, starting from some initial random PEPS state (in our analysis, we employ the $D=1$ random product state as the initial ansatz).  

It is expected that the imaginary time evolution converges to the best approximation of the ground state of the Hamiltonian. Unfortunately, this imaginary time simulation cannot be performed exactly for several reasons. First, the simulation relies on the Suzuki-Trotter decomposition of the evolution operator $\exp{[-H dt]}$ with a Trotter step $dt$. The finite value of the Trotter step introduces some errors into the final ground state approximation. To partially mitigate this source of errors, we first run the imaginary time evolution with a large Trotter step $dt=0.02$ until convergence, and then gradually decrease the Trotter step down to $dt=0.0002$. Second, the ground state usually cannot represent the ground state exactly due to the finite bond dimension~$D$. Because of this reason, the application of the Trotterized evolution operator to the PEPS state generally pushes the state away from the variational manifold, and the state should be projected back onto the manifold. To be more precise, the Trotterized gate application to the tensor network bond generally increases the bond dimension of the corresponding bond index. To return the state back to the manifold of states with the fixed maximal bond dimension, the grown bond index should be truncated back to the maximal bond dimension~$D$. In the Simple Update method, this truncation is not performed optimally, since it uses approximate bond environments based on belief propagation \cite{Jiang2008SimpleUpdate, TindallFishman2023GaugingBP}. An alternative approach using the full exact environments for bond truncations exists and is called Full Update \cite{Jordan2008iPEPS, Phien2015FastFullUpdate}, but we do not apply it here. We show in Sec.~\ref{sec:results} that corrections to Simple Update environments are small for the hyperhoneycomb lattice, and thus Simple Update should be rather accurate for the system under study. 

After the ground state is efficiently approximated with a tensor network state, one can compute some local observables for this state; the most interesting among these for us are the energies and the generalized local magnetizations. There are several different approaches how these observables can be evaluated for the PEPS states on 3d lattices. The first one was proposed in Ref.~\cite{JahromiOrus2019UniversalGPEPS} and uses Simple Update based environments (effectively, belief propagation based environments) to approximate local expectation values. This approach is astonishingly successful for gapped quantum systems but can be potentially unreliable for gapless systems. 
Its main drawback is that it lacks a control parameter for the observables accuracy and does not have a clear error estimate. The second approach was proposed in Ref.~\cite{VlaarCorboz2021Simulation3D} and later further developed in Refs.~\cite{LukinSotnikov2024SingleLayer3D,  LukinSotnikov2025SU4CubicDimerization,  Dziarmaga2026MonteCarlo3DPEPS}. This approach employs an accurate tensor-network-based contraction of the full sites environments with a combination of the boundary auxiliary PEPS states and of the corner transfer matrix renormalization group (CTMRG). Unfortunately, the computational cost of the method remains rather high, and it was not yet possible to reach bond dimensions $D>7$ with it. Yet another recently proposed approach suggests Monte-Carlo sampling from the PEPS state on the finite lattice \cite{Dziarmaga2026MonteCarlo3DPEPS}. However, in our current study, we work directly with the infinite lattice and thus, the approach is not applicable here. 

Finally, we should note that several related approaches were recently proposed based on loop or cluster expansions \cite{EvenblyPancottiMilstedGrayChan2026LoopSeries, GrayParkEvenblyPancottiKjonstadChan2025LoopClusterExpansions, MidhaZhang2025BeyondBP, MidhaSommersTindallAbanin2026RigorousBP}. These approaches are built upon the first mentioned approach \cite{JahromiOrus2019UniversalGPEPS, TindallFishman2023GaugingBP}, i.e., Simple Update or belief propagation environments, and controllably correct belief propagation based results for observables and partition functions with expansions in local clusters or loops around the sites of interest. Note that in this paper, we focus on the loop expansion variant, as defined in Ref.~\cite{EvenblyPancottiMilstedGrayChan2026LoopSeries}, and use the loop cluster expansion from Ref.~\cite{GrayParkEvenblyPancottiKjonstadChan2025LoopClusterExpansions} only in Appendix~\ref{appendix:comparison}. For a detailed exposition of the method, we refer to the study~\cite{EvenblyPancottiMilstedGrayChan2026LoopSeries} and only show the generalized loops relevant to the hyperhoneycomb lattice in Fig.~\ref{fig:loop_expansion}(c)--(g). We also discuss some issues with this loop expansion in more detail in Appendix~\ref{appendix:loop}. 
\begin{figure*}
    %\centering
    \includegraphics[width=\linewidth]{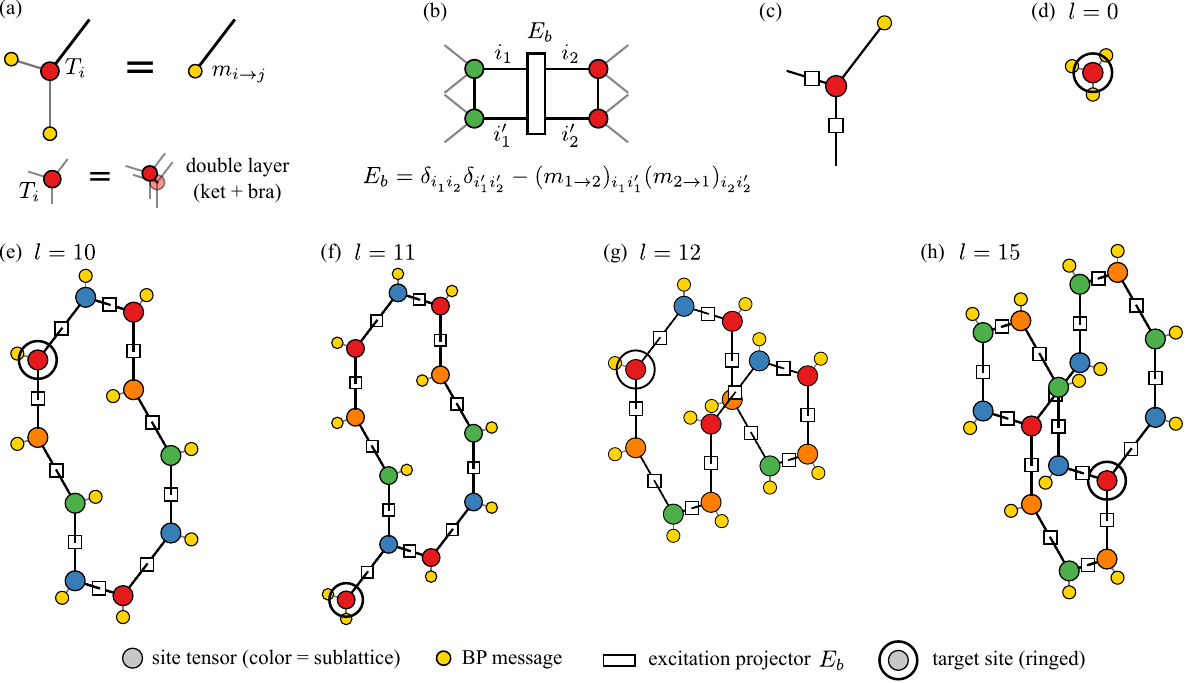}
    %{loop_expansion_hyperhoneycomb.pdf}
    \caption{The loop expansion on the hyperhoneycomb lattice. (a) We form the double layer tensors~$T_i$ from bra- and ket-state tensors and obtain belief propagation (BP) messages~$m_{i\to j}$ with iterative solution of BP equations \cite{TindallFishman2023GaugingBP}. (b) Definition of the ``excitation projectors" $E_b$, which are inserted on the bonds of loops formed by two BP messages~$m_{1\to2}$ and $m_{2\to1}$ on the bond \cite{EvenblyPancottiMilstedGrayChan2026LoopSeries}. (c) The ``transfer matrix" of the loop is a main building block of loop expansion correction. In case when all the tensors are approximately identical and isotropic, the leading eigenvalue $\lambda$ of this transfer matrix can be taken as the expansion parameter \cite{EvenblyPancottiMilstedGrayChan2026LoopSeries}. (d) The trivial BP contribution to the reduced density matrix (RDM) on the chosen target site (physical index between bra and ket on this site remains uncontracted). (e) The first nontrivial correction on the hyperhoneycomb lattice is the loop of the length 10. In all bonds internal to the loop, the projector is inserted, while all external bonds have BP messages on them. (f) Tadpole-like contribution to the RDM appearing at the loop order $l=11$. (g) A loop with $l=12$. (h) The first two-loop contribution appearing at loop order $l=15$.}
    \label{fig:loop_expansion}
\end{figure*}

The main control parameter of the loop expansions is the maximal loop order $l$---the number of sites in the loop. It was argued in Refs.~\cite{MidhaZhang2025BeyondBP, MidhaSommersTindallAbanin2026RigorousBP, EvenblyPancottiMilstedGrayChan2026LoopSeries} that the loop expansion generally converges exponentially with $l$. In particular, the order $l$ contribution is expected to scale as $\lambda^{l}$, where $\lambda$ is defined as a leading eigenvalue of the matrix shown in Fig.~\ref{fig:loop_expansion}(c). In Ref.~\cite{MidhaSommersTindallAbanin2026RigorousBP}, it was further shown that the magnitude of $\lambda$ depends on whether the target PEPS state is gapless or gapped. In particular, for gapped states, we can expect swift convergence of loop and cluster expansions, while for a gapless regime, the expansion quickly becomes unreliable. 

Here, we study the model with a potential gapless spin-liquid ground state; thus, one may expect that this approach becomes inapplicable to the problem at hand. In reality, this is not the case for two different reasons. First, the hyperhoneycomb lattice has very large loops of a minimal size $l_{\min}=10$. This is rather different from cubic lattices studied previously, which have minimal loop sizes $l_{\min}=4$. Hence, the leading correction to the local observables is expected to scale at least as $\lambda^{10}$, and corrections, at least up to the order $\lambda^{15}$, are still computable in practice, even for bond dimensions as high as $D=24$. The second reason is the nature of convergence of finite-$D$ PEPS states to the gapless infinite-$D$ limit. While in some applications (and with the help of advanced optimization methods, such as variational gradient optimization \cite{Corboz2016VariationalOpt, Vanderstraeten2016GradientMethods, LukinSotnikov2023HoneycombCTM, YangCorboz2026QRHoneycomb}), it is possible to obtain a gapless PEPS state already at a finite bond dimension $D$; in most cases, the finite-$D$ approximations to the true gapless spin-liquid ground state still have a finite correlation length \cite{Corboz2012SpinOrbitalLiquid, Liao2017GaplessKagome}, which scales to zero only in the limit $D=\infty$. This gives us hope that for small bond dimensions~$D$, the loop and cluster expansions can still be reliable, while the true gapless ground state can be studied with proper extrapolations. 

In our numerical analysis, we employ the tensor network \textsc{Julia} library \textsc{ITensors}~\cite{FishmanWhiteStoudenmire2022ITensor, FishmanWhiteStoudenmire2022ITensorRelease}. We also apply the \textsc{Python} package~\textsc{quimb}~\cite{Gray2018quimb} in calculations related to the loop and loop cluster expansions, in particular, for the purpose of combinatorial cluster generation.

\section{Results}\label{sec:results}

\subsection{Phases at finite $D$}\label{subsec:phases}
We begin our analysis of the model~\eqref{eq:model} by optimizing the wave functions for several different unit cells. Note that all unit cells should be multiples of the elementary unit cell consisting of four sites. The main cases under study are the $2 \times 2 \times 1$ and $1 \times 1 \times 1$ unit cells. Larger unit cells, such as $2 \times 2 \times 2$, are studied later in a separate subsection. 

The optimizations start from the bond dimension $D = 6$. While we can obtain results at smaller $D$, these have a much less pronounced structure of the phases. This is largely due to incomplete bond spectra multiplets for smaller bond dimensions. The optimizations for the chosen unit cells converge to three different phases specified below in the following paragraphs. 

\begin{figure}
    %\centering
    \includegraphics[width=\linewidth]{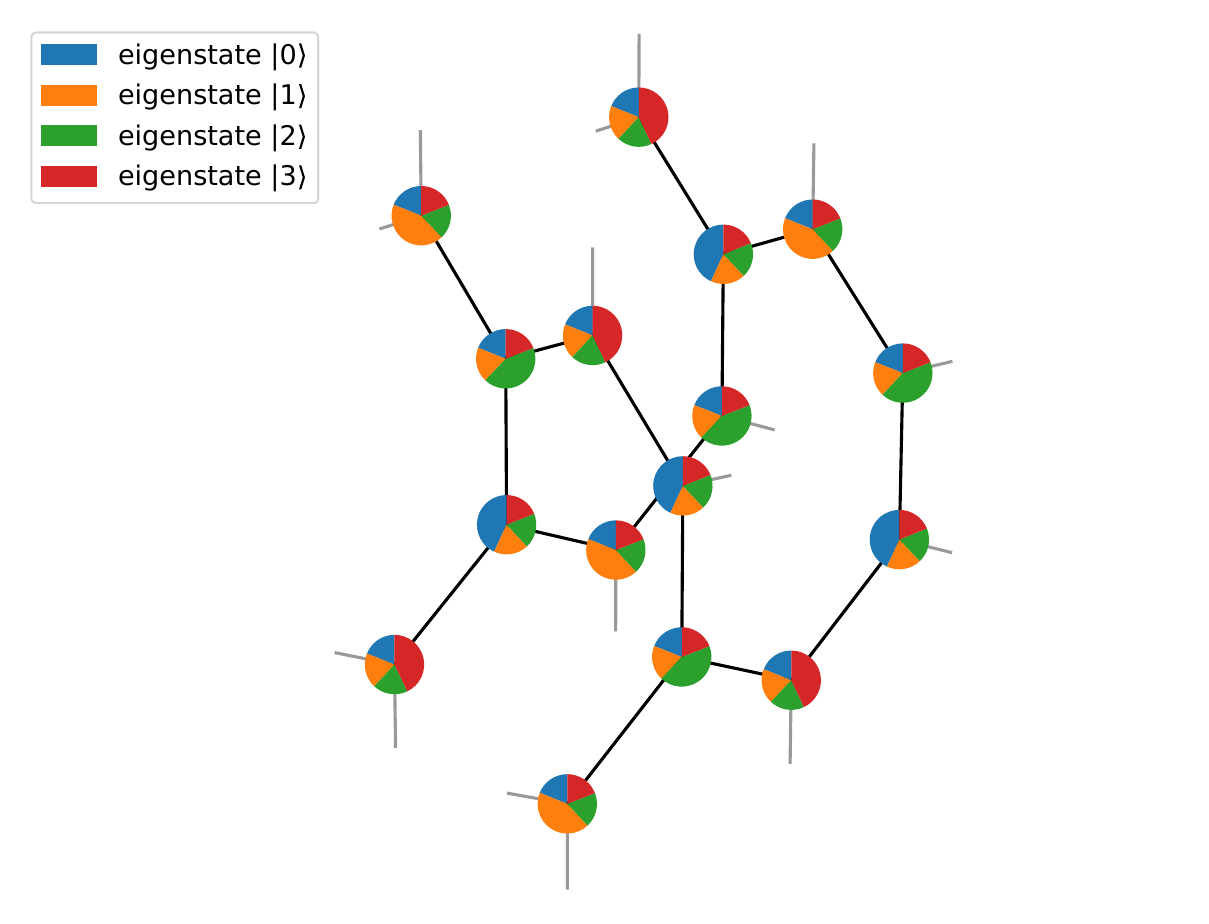}%{hyperhoneycomb_phases_magnetic_v2_D24.pdf}
    \caption{Magnetic phase with the $2 \times 2 \times 1$ unit cell. The values of densities of different color components are obtained at $l=15$ and $ D=24$.}
    \label{fig:2}
\end{figure}

The optimization with the $2 \times 2 \times 1$ unit cell mostly converges to the ``magnetic" phase with four different colors concentrated on four different sublattices of the hyperhoneycomb lattice. This phase is illustrated in Fig.~\ref{fig:2}, where one can see each site occupied by one color component (eigenstate~$|\alpha\rangle$) more than by the other three.

\begin{figure}
    %\centering
    \includegraphics[width=\linewidth]{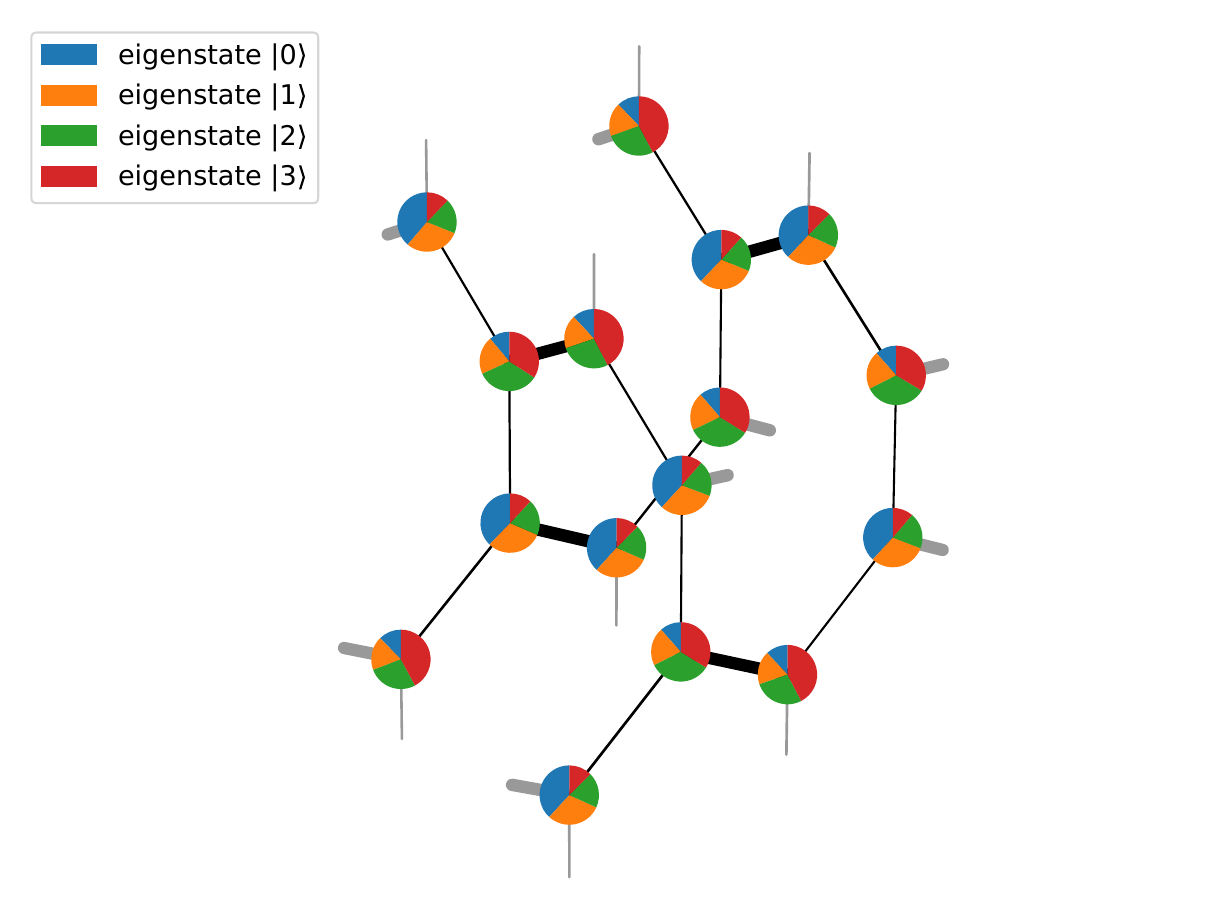}%{hyperhoneycomb_bonds_vbs_1_D6.pdf}
    \caption{$xy$-VBS phase with dimers (bold lines) on the $xy$ bonds. The width of the graph edges reflects the energy on this bond. The results are shown for the $2\times2\times1$ unit cell with $l=15$ and $ D=6$. The energies on dimerized and non-dimerized bonds are $E_d\approx-0.816$ and $E_{nd}\approx-0.399$, respectively.}
    \label{fig:3}
\end{figure}

In some cases, the optimization can also converge to the $xy$-dimerized (or, equivalently, $xy$-valence-bond-solid, $xy$-VBS) phase, which is illustrated in Fig.~\ref{fig:3}. This phase is characterized by a structure where two colors form a dimer placed on certain lattice bonds in $xy$ directions, while the two remaining colors form analogous dimers on the other bonds in $xy$ planes. This phase seems to be unstable at high bond dimensions. In particular, our strategy for obtaining this phase was as follows: first, optimize the initial PEPS wave function at bond dimension $D=4$ (where the wave function develops large anisotropy) and then use this wave function as the starting point for further optimization with a gradual increase of bond dimension to $D=5,6,7,8$. For yet higher bond dimensions, the wave function obtained in this way converges to the ``magnetic" phase. 

\begin{figure}
    %\centering
    \includegraphics[width=\linewidth]{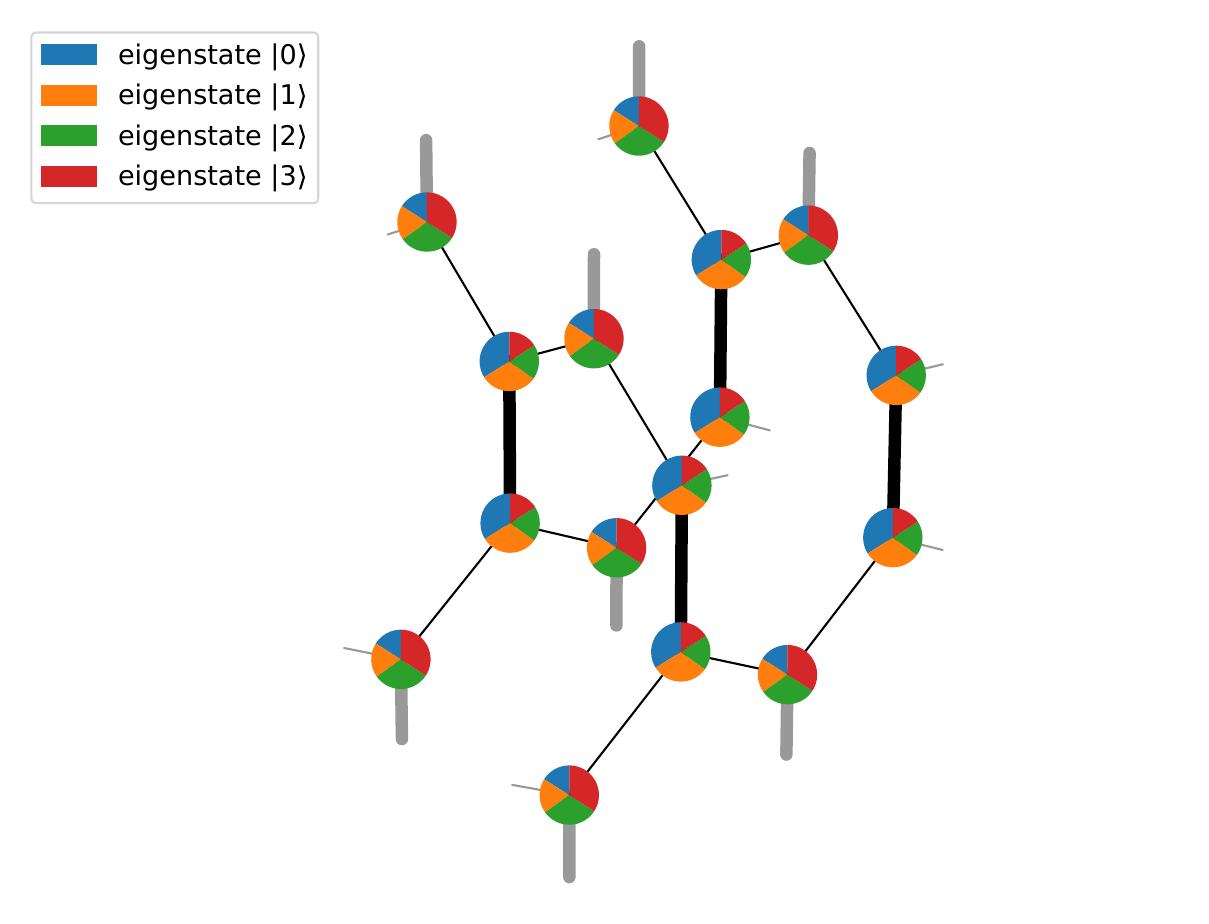}%{hyperhoneycomb_bonds_vbs_2_D24.pdf}
    \caption{$z$-VBS phase with dimers (bold lines) on the $z$ bonds. The width of the graph edges indicates the energy on this bond. The results are shown for the $2\times2\times1$ unit cell with $l=15$ and $D=24$. The energies are $E_d\approx-0.704$ and $E_{nd}\approx-0.520$.}
    \label{fig:4}
\end{figure}

The third phase, which we call $z$-dimerized ($z$-VBS) is shown in Fig. \ref{fig:4}  and is characterized by two-color dimers (valence bonds) oriented along the $z$ direction. At yet larger unit cells, it is possible to obtain ``spiral" magnetic phases, which are discussed in more detail in  Sec.~\ref{subsec:spiral}. 

\subsection{Observables}\label{subsec:observables}
The identified phases have different characteristic observables, which can be used for extrapolations to the infinite-$D$ limit. For the magnetic phase (as well as for the ``spiral" magnetic phases), we use the following magnetization-type observable: 
\begin{equation}
    m = \sum_{i} |3 \lambda_{i, 0} - \lambda_{i, 1} - \lambda_{i, 2} - \lambda_{i, 3}|/(3 N_{s}),
\end{equation}
where $i$ runs over all sites in the unit cell, $N_{s}$ is the number of sites in the unit cell, and $\lambda_{i, k}$ are the eigenvalues of the reduced density matrix (RDM) on site $i$ (where $\lambda_{i, 0}$ is the largest eigenvalue). Note that for the magnetic phase, we generally observe that $\lambda_{i, 1} \approx \lambda_{i,2} \approx \lambda_{i,3}$. For the unit RDM (peculiar to the featureless spin-liquid state), this observable is zero, $m = 0$. Hence, we can extrapolate this observable to the limit $D=\infty$ and check whether it approaches zero. If it does, then this is a clear indication of the spin-liquid state. 

For dimerized phases, we can define two different observables that characterize the dimerization strength. The first observable $d$ is defined as follows:
\begin{equation}\label{eq:dimeriz}
    d = \sum_{i} | \lambda_{i, 0} + \lambda_{i, 1} - \lambda_{i, 2} - \lambda_{i, 3}|/(2 N_{s}),
\end{equation}
where the quantities in the expression are defined in the same way as previously for $m$. For the dimerized state we generally expect $\lambda_{i, 0} \approx \lambda_{i,1}$ and $\lambda_{i,2} \approx \lambda_{i,3}$. 
Therefore, the observable $d$ effectively describes how much the local densities of dimerized colors are larger than the densities of non-dimerized colors.

Another important observable is the energy difference,
\begin{equation}
    \Delta E = \sum_{\langle ij\rangle_{nd} }\langle P_{ij} \rangle\bigr/N_{\langle i j \rangle_{nd}}
     - \sum_{\langle i j \rangle_{d}} \langle P_{ij} \rangle/N_{\langle i j \rangle_{d}}, 
\end{equation}
where the first sum is taken over expectation values of the permutation operator $P_{ij}$ on dimerized bonds~$\langle ij\rangle_{d}$ in the unit cell, while the second sum is taken over non-dimerized bonds~$\langle ij\rangle_{nd}$, and $N_{\langle i j \rangle_{d}}, N_{\langle i j \rangle_{nd}}$ are the numbers of dimerized and non-dimerized bonds in the unit cell. This observable counts the average energy difference between dimerized and non-dimerized bonds. Both $d$ and $\Delta E$ should extrapolate to zero for the featureless spin-liquid state.  

\begin{figure}
    %\centering
    \includegraphics[width=\linewidth]{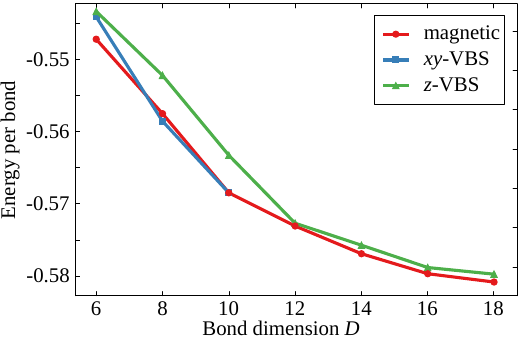}%{hyperhoneycomb_energy.pdf}
    \caption{Dependence of the average energy per bond on the bond dimension $D$ for three different possible phases. The results are obtained at the loop order $l=15$.}
    \label{fig:5}
\end{figure}

The last relevant observable, common to all phases, is the expectation value of the energy per bond. By comparing these observables in different phases, it is possible to determine the ground state at the given bond dimension~$D$. 
In particular, in Fig.~\ref{fig:5} we show the dependence of this quantity on the bond dimension for three different phases. The first observation is that the energy of the $z$-VBS phase is higher than the energy of the two other phases. This is consistent with the fact that the $z$-VBS phase converges only on the reduced size unit cell, while for larger unit cells, the PEPS state converges to the two other phases. The second observation is that the energies of the $xy$-VBS and the ``magnetic" phase become equal at the bond dimension $D=10$. This is actually due to the fact that the corresponding wave functions are also equal: the Simple Update optimization of the wave function initialized from the $D=8$ $xy$-VBS solution converges to the magnetic one at $D=10$. This is the reason we do not show the results for the $xy$-VBS phase at bond dimensions larger than $D=10$. The third observation is that the total energy shows significant variations depending on the bond dimension (even at comparably large bond dimensions), which points to a strongly correlated (spin-liquid) ground state. 

\begin{figure}
    %\centering
    \includegraphics[width=\linewidth]{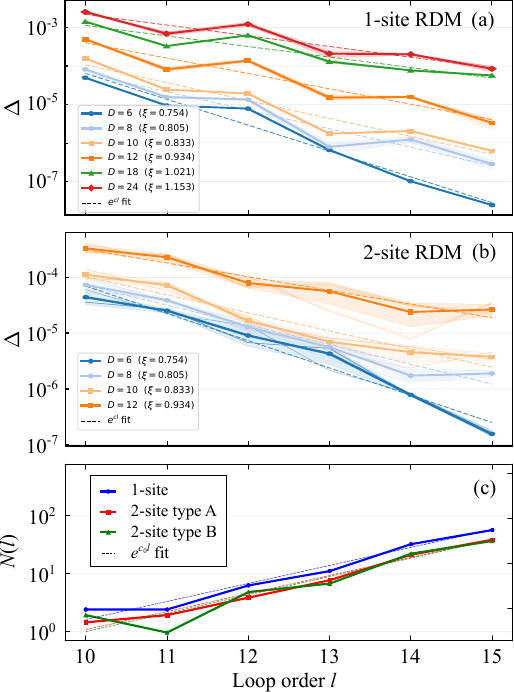}
    \caption{Convergence of the loop expansions of 1-site and 2-site local reduced density matrices. (a)--(b) Frobenius norm of the difference of the density matrices obtained at the consecutive loop orders, $\Delta=\langle||\rho(l)-\rho(l-1)||_F\rangle$. (c) Growth of different clusters $N(l)$ with cluster size $l$ for different RDMs. Note that there are two inequivalent types of 2-site bonds (type A and type B, where type B corresponds to $z$-directed bonds), which have different cluster counts. This increase can be approximated by exponential fits $N(l)\propto e^{c_0l}$ with $c_0 \approx 0.7$.  }
    \label{fig:loop_convergence}
\end{figure}

At this point, we should also discuss the convergence of the observables with the loop order. As we show in Fig.~\ref{fig:loop_convergence}, convergence is very fast at small bond dimensions~$D$, but becomes worse at larger $D$. The reason for this behavior is that at large bond dimensions, the state converges to the gapless phase and the correlation length diverges. At larger $\xi$, the loop corrections have much higher relative contributions \cite{MidhaSommersTindallAbanin2026RigorousBP}. 

Still, there are different convergence thresholds for different observables. For example, the ground state energy quickly converges with the bond dimension~$D$, and the differences in energy between nearest bond dimensions become small. For reliable results, the energy difference between different bond dimensions should still be larger than the error of loop expansion. Thus, we can compute the total energy reliably only for relatively small bond dimensions (in particular, in Fig.~\ref{fig:5} we show the results for the total energy up to $D=18$). On the other hand, physical observables such as dimerization and magnetization show relatively large differences (of the order of 1-2\%) between consecutive bond dimensions, even at relatively large $D \approx 20$. Hence, we can allow for much larger loop error for these observables, and our plots show these observables up to $D=24$. All the results shown below are computed with the maximal loop order $l=15$.

Note that our observations point toward exponential convergence of the reduced density matrices with loop order (which is confirmed by exponential fits in Fig.~\ref{fig:loop_convergence}). The number of clusters $N(l)$ also grows exponentially  with the loop order $l$; in particular, on the hyperhoneycomb lattice, the exponent of its growth is approximately $c_{0} = 0.7$, as illustrated in Fig.~\ref{fig:loop_convergence}(c). Hence, on general grounds, we can expect that the individual cluster contributions decay exponentially with some bond-dimension-dependent exponent, while the number of clusters grows exponentially with exponent $c_{0}$. The loop expansion fails to converge when these two exponents become equal. 

\subsection{Extrapolations to $D=\infty$}\label{subsec:extrapolation}
Before discussing the quantitative behavior of other observables, let us describe the possible methods of their extrapolation to the limit $D=\infty$. The first way to perform extrapolation is to postulate some ansatz, with a power-law behavior in bond dimension, for the observable:
\begin{equation}
    m(D) = m(\infty) + \frac{b}{D^{\beta}}.
\end{equation}
While this ansatz is widely used in the literature \cite{Liao2017GaplessKagome, Corboz2012SimplexSolids, Bauer2012ThreeSublatticeSU3}, we are not aware of any rigorous arguments as to why it should be valid. 

The second approach to the extrapolation to the infinite bond dimension is based on behavior with respect to the inverse of the correlation length~$\xi$ \cite{Corboz2018FCLS, RaderLauchli2018FCLS}, where the extrapolation ansatz has the form:
\begin{equation}
    m(D) = m(\infty) + \frac{b}{\xi(D)},
\end{equation}
where $\xi(D)$ is the correlation length of the state at bond dimension $D$. For traditional PEPS calculations, the correlation length $\xi(D)$ can be readily computed using transfer matrix constructed from the corner transfer matrix renormalization group (CTMRG) or boundary matrix product state (bMPS) tensors. Since we obtain all observables by using loop expansions, the correlation length is an observable, which is not directly accessed. In this study, we instead compute the ``tree" approximation of the correlation length, which is the correlation length of the correlation function computed with the leading approximation in the loop expansion. At small bond dimensions, where the loop expansion quickly converges according to our observations, this approximation of $\xi$ can be rather close to the correct one. Unfortunately, we cannot be sure that this approximation remains accurate at large bond dimensions. Additionally, there is no rigorous proof that the tree approximation of the correlation length limiting value at large bond dimension is infinite, as it is possible that the finite value of the tree correlation length together with proliferation of loop corrections leads to the infinite exact correlation length.

\begin{figure}
    %\centering
    \includegraphics[width=\linewidth]{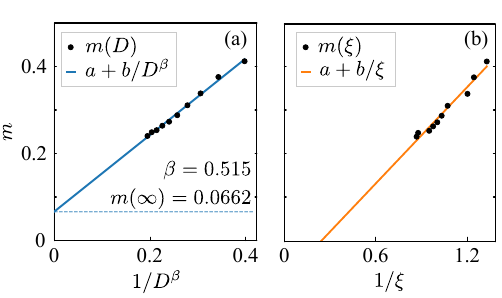}%{magnetizations_hyperhoneycomb.pdf}
    \caption{Extrapolation of the magnetization $m$ in the magnetic phase to the limit $D \to \infty$: (a) Extrapolation in $1/D^\beta$; (b) extrapolation in the inverse correlation length~$1/\xi$. }
    \label{fig:6}
\end{figure}

In Fig.~\ref{fig:6} we show the dependence of the magnetization $m$ on the bond dimension and inverse correlation length. The direct extrapolation in bond dimension leads to $m(\infty) \approx 0.066$. In terms of the onsite densities of components, this corresponds to $[0.28, 0.24, 0.24, 0.24]$, which is rather close to the symmetric case. Presumably, a more accurate extrapolation in correlation length predicts zero magnetization. From the given extrapolations, we can argue that the spin-flavor density distribution should be approximately uniform, which points to the gapless spin-liquid ground state. 

\begin{figure}
    %\centering
    \includegraphics[width=\linewidth]{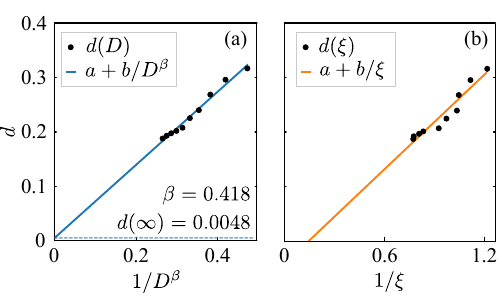}%{dimerizations_hyperhoneycomb.pdf}
    \caption{Extrapolation of the dimerization strength $d$ in the $z$-VBS state to the limit $D \to \infty$: (a) Extrapolation in $1/D^\beta$; (b) extrapolation in the inverse correlation length~$1/\xi$. }
    \label{fig:7}
\end{figure}

In Fig.~\ref{fig:7} we show the extrapolation of the dimerization~$d$, as determined in Eq.~\eqref{eq:dimeriz}. The direct fitting in bond dimension leads to nearly zero result, while the fitting in inverse correlation length further confirms this finding. Based on these fits, we can argue that the $z$-VBS state also decays to the featureless spin liquid in the large $D$ limit. 

\begin{figure}
    %\centering
    \includegraphics[width=\linewidth]{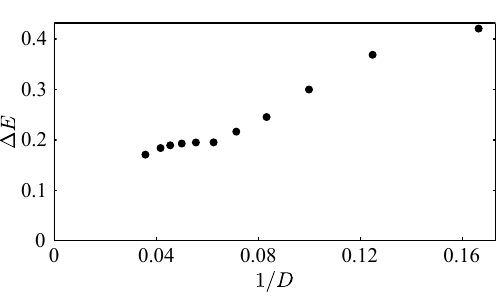}%{energy_anisotropy_hyperhoneycomb.pdf}
    \caption{Dependence of the energy anisotropy $\Delta E$ on the bond dimension $D$.  }
    \label{fig:8}
\end{figure}

Finally, in Fig.~\ref{fig:8} we show the dependence of the energy anisotropy $\Delta E$ in the $z$-VBS phase on the bond dimension $D$. In contrast to other observables, the energy anisotropy does not show any obvious fitting ansatz to extrapolate it, as the behavior at bond dimension $D<16$ is different from the behavior at $D \geq 16$. Note that the last point in Fig.~\ref{fig:8} corresponds to $D=28$ with the loop order $l=13$. This is the only place in our study where we have added this not fully converged result. The purpose is to show that the energy anisotropy continues to decay at larger bond dimensions. 

\subsection{Spiral states}\label{subsec:spiral}
Aside from the $2 \times 2 \times 1$ and $1 \times 1 \times 1$ unit cells described above, we have also studied the model on the larger unit cells, such as $4\times 4 \times 1$, $2 \times 2 \times 2$, and $2 \times 2 \times 4$. On these unit cells, we observe that Simple Update generally converges to ``magnetic" states, where a certain color dominates on every site of the lattice, but the precise pattern of magnetic sites can differ from the one found in the $2 \times 2 \times 1$ unit cell. In particular, the ordering can be non-collinear, when all the 1-site reduced density matrices cannot be simultaneously diagonalized. At the same time, observables like magnetization $m$ (as defined in the previous section) and total energy are nearly the same among these states.  Moreover, these states show the same convergence of magnetization to zero with larger bond dimensions. The reason for this behavior is that on the mean-field level, the model is highly degenerate: the mean-field equations force the nearest sites to be in orthogonal states, but there are many ways to fulfill these conditions, including those with non-collinear rotations. Some of these states are still stabilized by the Simple Update and have quite close energies even at high bond dimensions. 
Since we are not able to clearly exhaust all these variants of unit cells, we choose a different strategy: we obtain the states for several relatively small unit cells and also the ``spiral" states as described below, measure their energy and magnetization, and show that it is approximately the same as the magnetization and energy of the ``magnetic'' state on the $2 \times 2 \times 1$ unit cell.
We also show that the magnetization of these states decays to zero, and, therefore, all these states converge to the spin-liquid ground state at large bond dimensions. 

Let us now introduce the spiral states. The general idea follows the spiral iPEPS ansatz from Ref.~\cite{HasikCorboz2024SpiralPEPS} (see also Ref.~\cite{UedaMaruyama2012IncommensuratesMPS} as a direct MPS precursor). The spiral iPEPS state consists of a regular iPEPS state with a certain fixed unit cell, and the site-factorized and site-dependent unitary matrix $U({\bf q}) = \prod_{i} U_{i}({\bf q} \cdot {\bf r}_{i})$, where ${\bf q}$ is the spiral wavevector and $U_{i}$ acts only on site $i$. The matrix $U_{i} = \exp{[-i2\pi ({\bf q} \cdot {\bf r}_{i})K]}$, where $K$ is the Hermitian generator of the spiral rotation, which takes values in the Lie algebra of the model continuous symmetry group (in our case, in the SU(4) Lie algebra). If the local basis on site $i$ is spanned by states $|0\rangle, |1\rangle, |2\rangle, |3 \rangle$, then we choose the generator $K = i(|0\rangle \langle 1| - |1 \rangle \langle 0| + |2 \rangle \langle 3| -|3 \rangle \langle 2|)/2$. This form of generator is useful as it generates real-valued spiral unitaries and has a simple physical interpretation of spiral rotation between $|0\rangle$ and $|1\rangle$, as well as between $|2\rangle$ and $|3\rangle$. Note that more general spiral generators are possible as well, but we restrict ourselves to the above-specified generators. 

We chose the wavevector ${\bf q}$ to be oriented along the $z$-axis (along the vector of the inverse lattice, which is orthogonal to the direct lattice $z$ translation of the unit cell), and we use the notation $qz$ for its $z$ component. This choice of the vector ${\bf q}$ generally aligns with the magnetic orders, which were obtained for finite unit cells. Still, this restriction can be considered a limitation of our study. In the original spiral iPEPS paper, the spiral wavevector ${\bf q}$ was one of the variational parameters. In this study, we instead used a finite grid of ${\bf q}$ and separately optimized the iPEPS states with Simple Update for these values of ${\bf q}$. Note that spiral local unitaries lead to redefinition of the local Hamiltonian \cite{HasikCorboz2024SpiralPEPS}, but the Hamiltonian remains translationally invariant (with respect to the chosen unit cell), and the ground state can still be obtained with the Simple Update. 

\begin{figure}
    %\centering
    \includegraphics[width=\linewidth]{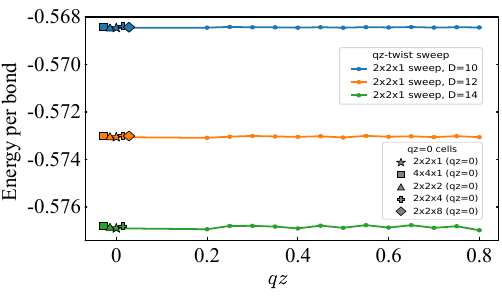}%{E_vs_qz.pdf}
    \caption{Energy $E$ averaged over bonds of magnetic states obtained for different spiral wavevectors $qz$ and for different finite unit cells.   }
    \label{fig:Energy_qz}
\end{figure}

\begin{figure}
    %\centering
    \includegraphics[width=\linewidth]{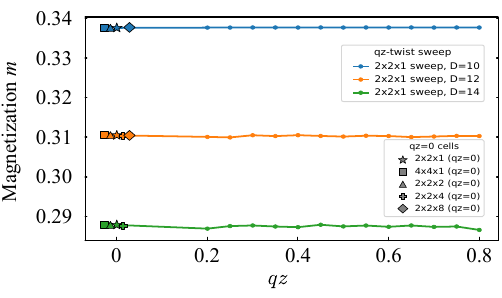}%{m_vs_qz.pdf}
    \caption{Magnetization $m$ of magnetic states obtained for different spiral wavevectors $qz$ and for different finite unit cells.   }
    \label{fig:magnetization_qz}
\end{figure}

In Fig.~\ref{fig:Energy_qz}, we show the energies at different bond dimensions ($D=10, 12, 14$) for magnetic states obtained with different spiral wavevectors or unit cells. The energies are very close to each other, and the differences can be caused by not fully converged effects in the Simple Update scheme.  In Fig.~\ref{fig:magnetization_qz}, we show the previously defined magnetization $m$ for the same states. These observables are again very close to each other. This means that they can be extrapolated together to the infinite bond dimension limit, and the result of extrapolation should be the same as the result for the simplest magnetic state on the $2 \times 2 \times 1$ unit cell discussed in the previous subsection. Therefore, all these states should converge to the spin-liquid state in the large $D$ limit. 

\section{Conclusions and Outlook}\label{sec:conclusion}

In this study, we analyzed the ground state of the SU(4) Heisenberg model on the hyperhoneycomb lattice using iPEPS variational ansatz. At finite bond dimensions, we were able to identify three different possible phases, which we call ``magnetic", ``$z$-VBS", and ``$xy$-VBS" phases. We employed recently introduced loop expansions on top of belief-propagation to compute observables, such as bond energies and local densities of different spin components (colors). At finite bond dimensions, the ``magnetic" phase generally shows lower energy. We further extrapolated the observables as local densities to the infinite bond dimension regime ($D \to \infty$) and obtained the approximately featureless uniform state. Together with a constant growth of the correlation length, this result points to the gapless spin-liquid ground state, which is in agreement with previous studies. 

Our main methodological claim is that for some ``tree-like" lattices with small connectivity and large loops, it is possible to use tensor networks together with the loop expansion to determine the ground state, even for the gapless case. The two necessary conditions for this method are small and swiftly decaying loop corrections (which are expected on tree-like lattices) and the finite correlation length of the optimized tensor network state due to the finite bond dimension. The second condition generally holds when the PEPS state is optimized with the Simple Update algorithm (although, for the variational-gradient-based optimization of iPEPS, it is possible to obtain an infinite correlation length at finite bond dimensions). We expect that the methodology of this study can be applied to other interesting models on the hyperhoneycomb lattice, in particular, to the Kitaev--Heisenberg model \cite{JahromiYarlooOrus2021Kitaev3D, LeeSchafferBhattacharjeeKim2014HKHyperhoneycomb} and thereby to the description of hyperhoneycomb iridates~\cite{TakayamaKatoDinnebierNusTakagi2015BetaLi2IrO3}. The method can potentially be applied to other tree-like lattices, such as hyperoctagon lattice, stripy-hyperhoneycomb lattice (with minimal loop length $14$), the Heisenberg model on the hyperkagome lattice \cite{OkamotoNoharaTakagi2007Hyperkagome, BergholtzLauchliMoessner2010Hyperkagome} (with the iPESS ansatz \cite{XieChenYuKongNormandXiang2014PESS} instead of the iPEPS ansatz of this study), and 2d heavy-hex lattice underlying IBM QPU. It may also be interesting to try applying the method to hyperbolic lattices, which are known to have mean-field-like phase transition, and observables on such lattices converge very quickly with CTMRG bond dimensions \cite{GendiarKrcmarAndergassenDaniskaNishino2012HyperbolicIsing, DaniskaGendiar2016QuantumSpinHyperbolic}.

\appendix
\section{Discussion of disconnected loops in loop expansion}\label{appendix:loop}

The loop expansion of a local reduced density matrix is based on the Belief propagation (BP) environments as a leading approximation. Hence, the (non-normalized) density matrix can be represented as $\rho = \rho_{BP} +\rho_{loops}$, where $\rho_{BP}$ represents the leading belief-propagation-based term and $\rho_{loops}$ are all the loop corrections. The loop corrections are sums over all connected and disconnected clusters up to one tadpole condition, which states that there cannot exist a vertex with only one incoming line (unless this is the vertex where the density matrix is defined). The presence of disconnected graphs naively poses a problem in evaluating the density matrices (as well as the state norm). The reason is that for a system of size $L$, we can expect order $L$ graphs with a single disconnected loop, order $L^{2}$ graphs with two different disconnected loops, etc. In the thermodynamic limit, these contributions naively diverge. 

For the norm calculation, the solution was found in Ref. \cite{MidhaZhang2025BeyondBP}, where loop expansion was generalized to the $\log \langle \psi| \psi\rangle$ computation, which is linear (extensive) in system size $L$. Then, $F = \log \langle \psi| \psi\rangle/L$ can be computed with the summation of only connected graphs. Now, the same approach can be used in the computation of the loop expansion for the reduced density matrices. First, we can group together all disconnected graphs with the same connected components: 
$\rho = \rho_{0}\times Z_{0} +\rho_{1}\times Z_{1} + \dots$, where $\rho_{0}$ is the contribution of the trivial connected cluster (reduced density matrix site tensors and incoming BP messages), $\rho_{1}$ is the first connected correction (in our case, the length $10$ loop on the hyperhoneycomb lattice, piercing through one of the reduced density matrix sites), $Z_{0}$ is a sum over all disconnected graphs with a trivial connected part, and $Z_{1}$ is a sum over all disconnected graphs with a 1-loop connected part. 

Equivalently, $Z_{0}$ and $Z_{1}$ can be represented as deformations of the tensor network norm contraction, where all the bonds connecting the trivial or 1-loop graph to the rest of the system are replaced by the BP messages coming to the rest of the system. Now, these quantities should have the following behavior with respect to system size $L$: $Z_{0} = \exp{[L F + F_{0}]}$, $Z_{1} = \exp{[L F + F_{1}]}$, where $F$ is some identical extensive part, while $F_{0}$ and $F_{1}$ are independent of $F$ boundary contributions, characterizing the connected graph that was cut off from the rest of the system by BP messages.  

Now, it is possible to finally normalize the reduced density matrix by the trace, stripping it from the infinite $\exp{[LF + F_{0}]}$ factor.  As a result, the reduced density matrix has the form $\rho = \rho_{0} + \rho_{1} \exp{[F_{1} - F_{0}]} + \cdots$. The important detail here is that $F_{1}$ and $F_{0}$ can also be systematically computed by the loop expansion, with the leading contribution corresponding to the length $10$ loops, which is of order $\lambda^{10}$ in the cluster size. Note that the contribution $\rho_{1}$ is already of the order $\lambda^{10}$ itself. Hence, the factor $\exp{[F_{1} - F_{0}]}$ can be neglected as long as the loop expansion stops at loop size smaller than $20$. This is the main reason why we have used just the loop expansion in connected loops to find the observables.

\section{Comparison of loop expansion and loop cluster expansion}\label{appendix:comparison}

Recently, Refs. \cite{ParkGrayChan2025InfluenceFunctionalBP, GrayParkEvenblyPancottiKjonstadChan2025LoopClusterExpansions} proposed a different version of expansions on top of BP in different clusters---loop cluster expansion. Both methods are approximately equivalent in their computational cost (as they contract nearly identical tensor networks), but they combine the cluster contributions in different ways. 

\begin{figure}
    %\centering
    \includegraphics[width=\linewidth]{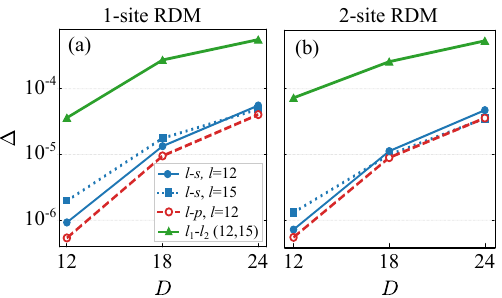}
    %{compare_loop_cluster_diff.pdf}
    \caption{Comparison of the density matrices in terms of the maximal Frobenius norm of the RDM difference along the unit cell, $\Delta=\max||\rho_a -\rho_b||$, between loop expansions and loop cluster expansions (sum and product formulas separately) at different bond dimensions~$D$ and loop orders~$l$. For comparison, we also show the difference~$\Delta$ between RDM at loop expansions $l_1=12$ and $l_2=15$.  }
    \label{fig:cluster}
\end{figure}

Here, let us employ the loop cluster expansions to crosscheck loop expansions and to confirm that both methods converge to the same density matrix. 
In Fig.~\ref{fig:cluster} we show a comparison of the loop expansions with loop cluster expansions (with ``sum" and ``product" formulas from Ref.~\cite{GrayParkEvenblyPancottiKjonstadChan2025LoopClusterExpansions}). For comparison, we also compare the reduced density matrices obtained with loop expansion at loop orders $l=12$ and $l=15$. Note that we do not show the results obtained with the loop cluster expansion ``product" formula at $l=15$, as this quantity becomes singular in certain cases. There are two clear observations about the two methods: the difference between the two methods is rather small (much smaller than the difference between different orders of loop expansion); the difference clearly increases with bond dimension, which is expected due to the larger correlation length (see Ref.\cite{MidhaSommersTindallAbanin2026RigorousBP}). 

\acknowledgements
The authors acknowledge support by the National Research Foundation of Ukraine, project No.~2023.03/0073.

\bibliography{refs}

\end{document}